%% file: main.tex
\pdfoutput=1

\documentclass[12pt,a4paper]{article}

\usepackage{ifthen} 
\newboolean{pdflatex}
\setboolean{pdflatex}{true} 

\newboolean{articletitles}
\setboolean{articletitles}{true} 

\newboolean{uprightparticles}
\setboolean{uprightparticles}{false} 

\newboolean{inbibliography}
\setboolean{inbibliography}{false} 

\input{preamble}
\usepackage{longtable} 

\begin{document}

\renewcommand{\thefootnote}{\fnsymbol{footnote}}
\setcounter{footnote}{1}

\input{title-LHCb-PAPER}


\renewcommand{\thefootnote}{\arabic{footnote}}
\setcounter{footnote}{0}



\pagestyle{plain} 
\setcounter{page}{1}
\pagenumbering{arabic}


%

\input{body}

\input{acknowledgements}



\addcontentsline{toc}{section}{References}
\setboolean{inbibliography}{true}

\ifx\mcitethebibliography\mciteundefinedmacro
\PackageError{LHCb.bst}{mciteplus.sty has not been loaded}
{This bibstyle requires the use of the mciteplus package.}\fi
\providecommand{\href}[2]{#2}


\end{document}

%% file: preamble.tex
\usepackage{lineno}  
\usepackage{xspace} 
\usepackage{caption} 

\usepackage{graphicx}  
\usepackage{color}
\usepackage{colortbl}
\graphicspath{{./figs/}} 

\usepackage{amsmath} 
\usepackage{amssymb}
\usepackage{amsfonts}
\usepackage{upgreek} 

\newcommand*\patchAmsMathEnvironmentForLineno[1]{%
\expandafter\let\csname old#1\expandafter\endcsname\csname #1\endcsname
\expandafter\let\csname oldend#1\expandafter\endcsname\csname
end#1\endcsname
 \renewenvironment{#1}%
   {\linenomath\csname old#1\endcsname}%
   {\csname oldend#1\endcsname\endlinenomath}%
}
\newcommand*\patchBothAmsMathEnvironmentsForLineno[1]{%
  \patchAmsMathEnvironmentForLineno{#1}%
  \patchAmsMathEnvironmentForLineno{#1*}%
}
\AtBeginDocument{%
\patchBothAmsMathEnvironmentsForLineno{equation}%
\patchBothAmsMathEnvironmentsForLineno{align}%
\patchBothAmsMathEnvironmentsForLineno{flalign}%
\patchBothAmsMathEnvironmentsForLineno{alignat}%
\patchBothAmsMathEnvironmentsForLineno{gather}%
\patchBothAmsMathEnvironmentsForLineno{multline}%
}

\usepackage{hyperref}    
\usepackage[all]{hypcap} 

\input{lhcb-symbols-def} 
\input{aliases}

\usepackage{cite} 
\usepackage{mciteplus}

%% file: lhcb-symbols-def.tex
\def\lhcb {\mbox{LHCb}\xspace}

\ifthenelse{\boolean{uprightparticles}}%
{

 \def\Ppi         {\ensuremath{\uppi}\xspace}

 \def\Ppsi        {\ensuremath{\uppsi}\xspace}

 \def\PDelta      {\ensuremath{\Delta}\xspace}                 
 \def\PXi      {\ensuremath{\Xi}\xspace}                 
 \def\PLambda      {\ensuremath{\Lambda}\xspace}                 
 \def\PSigma      {\ensuremath{\Sigma}\xspace}                 
 \def\POmega      {\ensuremath{\Omega}\xspace}                 
 \def\PUpsilon      {\ensuremath{\Upsilon}\xspace}                 
 
 \def\PB      {\ensuremath{\mathrm{B}}\xspace}                 
                  
 \def\PD      {\ensuremath{\mathrm{D}}\xspace}

 \def\PJ      {\ensuremath{\mathrm{J}}\xspace}                 
 \def\PK      {\ensuremath{\mathrm{K}}\xspace}

 \def\Pb      {\ensuremath{\mathrm{b}}\xspace}                 
 \def\Pc      {\ensuremath{\mathrm{c}}\xspace}

 \def\Pi      {\ensuremath{\mathrm{i}}\xspace}

 \def\Ps      {\ensuremath{\mathrm{s}}\xspace}

}
{

 \def\Ppi         {\ensuremath{\pi}\xspace}

 \def\Ppsi        {\ensuremath{\psi}\xspace}                 
                  
 \mathchardef\PDelta="7101
 \mathchardef\PXi="7104
 \mathchardef\PLambda="7103
 \mathchardef\PSigma="7106
 \mathchardef\POmega="710A
 \mathchardef\PUpsilon="7107
                  
 \def\PB      {\ensuremath{B}\xspace}                 
                  
 \def\PD      {\ensuremath{D}\xspace}

 \def\PJ      {\ensuremath{J}\xspace}                 
 \def\PK      {\ensuremath{K}\xspace}

 \def\Pb      {\ensuremath{b}\xspace}                 
 \def\Pc      {\ensuremath{c}\xspace}

 \def\Pi      {\ensuremath{i}\xspace}

 \def\Ps      {\ensuremath{s}\xspace}

}

\def\squark    {\ensuremath{\Ps}\xspace}

\def\cquark    {\ensuremath{\Pc}\xspace}
\def\cquarkbar {\ensuremath{\overline \cquark}\xspace}
\def\ccbar     {\ensuremath{\cquark\cquarkbar}\xspace}
\def\bquark    {\ensuremath{\Pb}\xspace}

\def\pion  {\ensuremath{\Ppi}\xspace}

\def\pip   {\ensuremath{\pion^+}\xspace}
\def\pim   {\ensuremath{\pion^-}\xspace}

\def\kaon  {\ensuremath{\PK}\xspace}
  \def\Kbar  {\kern 0.2em\overline{\kern -0.2em \PK}{}\xspace}

\def\Kp    {\ensuremath{\kaon^+}\xspace}
\def\Km    {\ensuremath{\kaon^-}\xspace}

\def\KS    {\ensuremath{\kaon^0_{\rm\scriptscriptstyle S}}\xspace}

  \def\Dbar    {\kern 0.2em\overline{\kern -0.2em \PD}{}\xspace}
\def\D       {\ensuremath{\PD}\xspace}

\def\Dz      {\ensuremath{\D^0}\xspace}
\def\Dzb     {\ensuremath{\Dbar^0}\xspace}
\def\Dp      {\ensuremath{\D^+}\xspace}
\def\Dm      {\ensuremath{\D^-}\xspace}

\def\Dstarp  {\ensuremath{\D^{*+}}\xspace}

\def\Dsp     {\ensuremath{\D^+_\squark}\xspace}
\def\Dsm     {\ensuremath{\D^-_\squark}\xspace}

\def\B       {\ensuremath{\PB}\xspace}
\def\Bbar    {\ensuremath{\kern 0.18em\overline{\kern -0.18em \PB}{}}\xspace}

\def\Bz      {\ensuremath{\B^0}\xspace}
\def\Bzb     {\ensuremath{\Bbar^0}\xspace}
\def\Bu      {\ensuremath{\B^+}\xspace}
\def\Bub     {\ensuremath{\B^-}\xspace}
\def\Bp      {\ensuremath{\Bu}\xspace}
\def\Bm      {\ensuremath{\Bub}\xspace}

\def\Bs      {\ensuremath{\B^0_\squark}\xspace}
\def\Bsb     {\ensuremath{\Bbar^0_\squark}\xspace}

\def\jpsi     {\ensuremath{{\PJ\mskip -3mu/\mskip -2mu\Ppsi\mskip 2mu}}\xspace}

  \def\Y#1S{\ensuremath{\PUpsilon{(#1S)}}\xspace}

\def\Lz {\ensuremath{\PLambda}\xspace}
\def\Lbar {\ensuremath{\kern 0.1em\overline{\kern -0.1em\PLambda}}\xspace}

\def\Lb      {\ensuremath{\Lz^0_\bquark}\xspace}

\def\Lc      {\ensuremath{\Lz^+_\cquark}\xspace}

\def\to                 {\ensuremath{\rightarrow}\xspace}

\def\CP                {\ensuremath{C\!P}\xspace}

\newcommand{\DGs}{\ensuremath{\Delta\Gamma_{\squark}}\xspace}

\newcommand{\Gs}{\ensuremath{\Gamma_{\squark}}\xspace}

\newcommand{\GL}{\ensuremath{\Gamma_{\rm L}}\xspace}
\newcommand{\GH}{\ensuremath{\Gamma_{\rm H}}\xspace}

\def\AT#1     {\ensuremath{A_{\mathrm{T}}^{#1}}\xspace}           

\def\C#1      {\ensuremath{\mathcal{C}_{#1}}\xspace}                       
\def\Cp#1     {\ensuremath{\mathcal{C}_{#1}^{'}}\xspace}                    
\def\Ceff#1   {\ensuremath{\mathcal{C}_{#1}^{\mathrm{(eff)}}}\xspace}        
\def\Cpeff#1  {\ensuremath{\mathcal{C}_{#1}^{'\mathrm{(eff)}}}\xspace}       
\def\Ope#1    {\ensuremath{\mathcal{O}_{#1}}\xspace}                       
\def\Opep#1   {\ensuremath{\mathcal{O}_{#1}^{'}}\xspace}                    

\newcommand{\tev}{\ifthenelse{\boolean{inbibliography}}{\ensuremath{~T\kern -0.05em eV}\xspace}{\ensuremath{\mathrm{\,Te\kern -0.1em V}}\xspace}}
\newcommand{\gev}{\ensuremath{\mathrm{\,Ge\kern -0.1em V}}\xspace}
\newcommand{\mev}{\ensuremath{\mathrm{\,Me\kern -0.1em V}}\xspace}
\newcommand{\kev}{\ensuremath{\mathrm{\,ke\kern -0.1em V}}\xspace}
\newcommand{\ev}{\ensuremath{\mathrm{\,e\kern -0.1em V}}\xspace}
\newcommand{\gevc}{\ensuremath{{\mathrm{\,Ge\kern -0.1em V\!/}c}}\xspace}
\newcommand{\mevc}{\ensuremath{{\mathrm{\,Me\kern -0.1em V\!/}c}}\xspace}
\newcommand{\gevcc}{\ensuremath{{\mathrm{\,Ge\kern -0.1em V\!/}c^2}}\xspace}
\newcommand{\gevgevcccc}{\ensuremath{{\mathrm{\,Ge\kern -0.1em V^2\!/}c^4}}\xspace}
\newcommand{\mevcc}{\ensuremath{{\mathrm{\,Me\kern -0.1em V\!/}c^2}}\xspace}

\def\mum  {\ensuremath{{\,\upmu\rm m}}\xspace}

\def\invfb   {\ensuremath{\mbox{\,fb}^{-1}}\xspace}

\def\ps   {\ensuremath{{\rm \,ps}}\xspace}

\newcommand{\stat}{\ensuremath{\mathrm{\,(stat)}}\xspace}
\newcommand{\syst}{\ensuremath{\mathrm{\,(syst)}}\xspace}

\newcommand{\chisq}{\ensuremath{\chi^2}\xspace}

\newcommand{\chisqip}{\ensuremath{\chi^2_{\rm IP}}\xspace}
\newcommand{\chisqvs}{\ensuremath{\chi^2_{\rm VS}}\xspace}
\newcommand{\chisqvtx}{\ensuremath{\chi^2_{\rm vtx}}\xspace}

\def\gsim{{~\raise.15em\hbox{$>$}\kern-.85em
          \lower.35em\hbox{$\sim$}~}\xspace}
\def\lsim{{~\raise.15em\hbox{$<$}\kern-.85em
          \lower.35em\hbox{$\sim$}~}\xspace}

\def\pt         {\mbox{$p_{\rm T}$}\xspace}

\def\evtgen     {\mbox{\textsc{EvtGen}}\xspace}

\def\gauss      {\mbox{\textsc{Gauss}}\xspace}
\def\geant      {\mbox{\textsc{Geant4}}\xspace}

\def\photos     {\mbox{\textsc{Photos}}\xspace}

\def\pythia     {\mbox{\textsc{Pythia}}\xspace}

\def\tell1  {TELL1\xspace}
\def\ukl1   {UKL1\xspace}



%% file: aliases.tex
\def\bstodspipipi{\Bsb\to D_s^+\pi^-\pi^+\pi^-}

\def\btodzds{\Bm\to\Dz\Dsm}

\def\bstodsds{\Bsb\to\Dsm\Dsp}

\def\bstodsd{\Bsb\to\Dm\Dsp}
\def\btodsd{\Bz\to\Dm\Dsp}

\def\btodzds{\Bm\to\Dz\Dsm}

\def\taueff{\tau^{\rm eff}_{\bstodsds}}
\def\taueffkk{\tau^{\rm eff}_{\Bsb\to\Kp\Km}}

%% file: title-LHCb-PAPER.tex

\begin{titlepage}
\pagenumbering{roman}

\vspace*{-1.5cm}
\centerline{\large EUROPEAN ORGANIZATION FOR NUCLEAR RESEARCH (CERN)}
\vspace*{1.5cm}
\hspace*{-0.5cm}
\begin{tabular*}{\linewidth}{lc@{\extracolsep{\fill}}r}
\ifthenelse{\boolean{pdflatex}}
{\vspace*{-2.7cm}\mbox{\!\!\!\includegraphics[width=.14\textwidth]{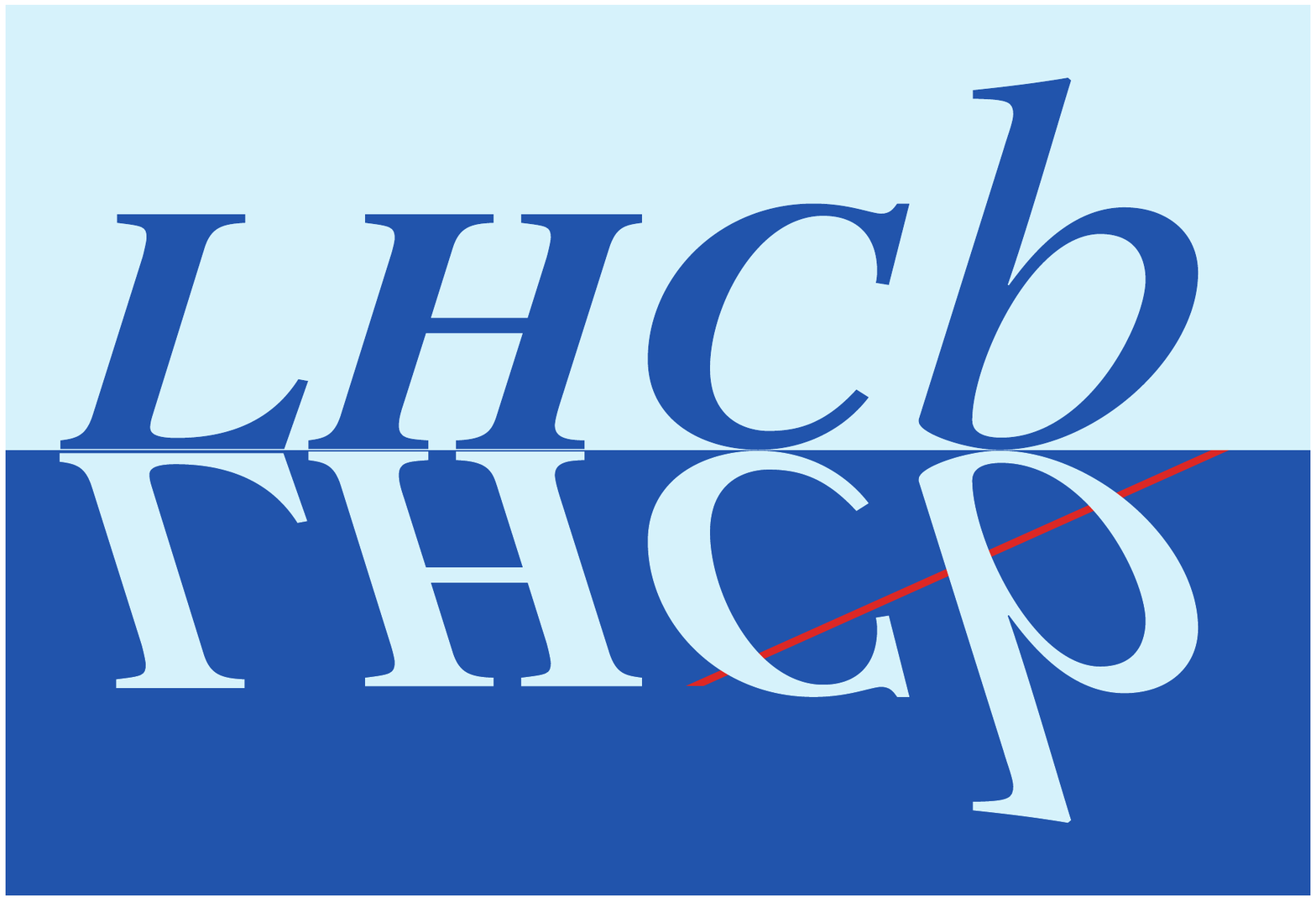}} & &}%
{\vspace*{-1.2cm}\mbox{\!\!\!\includegraphics[width=.12\textwidth]{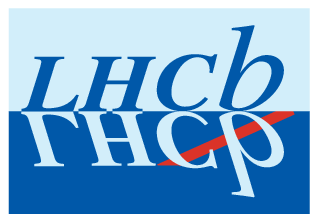}} & &}%
\\
 & & CERN-PH-EP-2013-218 \\  
 & & LHCb-PAPER-2013-060 \\  
 & & December 4, 2013 \\ 
 & & \\
\end{tabular*}

\vspace*{4.0cm}

{\bf\boldmath\huge
\begin{center}
  Measurement of the $\bstodsds$ and $\bstodsd$ effective lifetimes 
\end{center}
}

\vspace*{2.0cm}

\begin{center}
The LHCb collaboration\footnote{Authors are listed on the following pages.}
\end{center}

\vspace{\fill}

\begin{abstract}
  \noindent
The first measurement of the effective lifetime of the $\Bsb$ meson in the decay $\bstodsds$ is reported using
a proton-proton collision dataset, corresponding to an integrated luminosity of 3\invfb, collected by the LHCb experiment.
The measured value of the $\bstodsds$ effective lifetime is
$1.379\pm0.026\pm0.017$~ps, where the uncertainties are statistical and systematic, respectively. 
This lifetime translates into a measurement of the decay width of the light $\Bsb$
mass eigenstate of \GL$=0.725\pm0.014\pm0.009$~ps$^{-1}$. The $\Bsb$ lifetime is also
measured using the flavor-specific $\bstodsd$ decay to be $1.52\pm0.15\pm0.01~{\rm ps}$.
\end{abstract}

\vspace*{2.0cm}

\begin{center}
  Submitted to Phys.~Rev.~Lett. 
\end{center}

\vspace{\fill}

{\footnotesize 
\centerline{\copyright~CERN on behalf of the \lhcb collaboration, license \href{http://creativecommons.org/licenses/by/3.0/}{CC-BY-3.0}.}}
\vspace*{2mm}

\end{titlepage}


\newpage
\setcounter{page}{2}
\mbox{~}
\newpage

\input{LHCb_HD_authorlist_2013-10-27.tex}

\cleardoublepage

%% file: LHCb_HD_authorlist_2013-10-27.tex
\centerline{\large\bf LHCb collaboration}
\begin{flushleft}
\small
R.~Aaij$^{40}$, 
B.~Adeva$^{36}$, 
M.~Adinolfi$^{45}$, 
A.~Affolder$^{51}$, 
Z.~Ajaltouni$^{5}$, 
J.~Albrecht$^{9}$, 
F.~Alessio$^{37}$, 
M.~Alexander$^{50}$, 
S.~Ali$^{40}$, 
G.~Alkhazov$^{29}$
% ...

{\footnotesize \it
% ...
$ ^{5}$Clermont Universit\'{e}, Universit\'{e} Blaise Pascal, CNRS/IN2P3, LPC, Clermont-Ferrand, France\\
% ...
$ ^{9}$Fakult\"{a}t Physik, Technische Universit\"{a}t Dortmund, Dortmund, Germany\\
% ...
$ ^{29}$Petersburg Nuclear Physics Institute (PNPI), Gatchina, Russia\\
% ...
$ ^{36}$Universidad de Santiago de Compostela, Santiago de Compostela, Spain\\
$ ^{37}$European Organization for Nuclear Research (CERN), Geneva, Switzerland\\
% ...
$ ^{40}$Nikhef National Institute for Subatomic Physics, Amsterdam, The Netherlands\\
% ...
$ ^{45}$H.H. Wills Physics Laboratory, University of Bristol, Bristol, United Kingdom\\
% ...
$ ^{50}$School of Physics and Astronomy, University of Glasgow, Glasgow, United Kingdom\\
$ ^{51}$Oliver Lodge Laboratory, University of Liverpool, Liverpool, United Kingdom\\
% ...
}
\end{flushleft}

%% file: body.tex
A central goal in quark-flavor physics is to test whether
the Cabibbo-Kobayashi-Maskawa (CKM) mechanism~\cite{Cabibbo:1963yz,Kobayashi:1973fv} can fully describe all
relevant weak decay observables, or if physics beyond the Standard Model (SM) is needed. 
In the neutral $B$ meson sector, the mass eigenstates do not coincide with the flavor eigenstates as a result of
$B\Bbar$ mixing.
In addition to measurable mass splittings between the mass eigenstates~\cite{LHCb-PAPER-2013-006}, 
the $B_s$ system also exhibits a sizeable difference in the decay widths $\GL$ and $\GH$, where
the subscripts ${\rm L}$ and ${\rm H}$ refer to the light and heavy mass eigenstates, respectively. This difference
is due to the large decay width to final states accessible to both $\Bs$ and $\Bsb$. 
In the absence of \CP violation, the mass eigenstates are also eigenstates
of \CP. The summed decay rate of $\Bs$ and $\Bsb$ to the \CP-even $\Dsp\Dsm$ final state 
can be written as~\cite{Fleischer:2011cw}
\begin{align}
\label{eq:decayrate1}
\Gamma_{\bstodsds}(t) + \Gamma_{\Bs\to\Dsp\Dsm}(t) \propto (1 + \cos\phi_s)e^{-\GL t} + (1 - \cos\phi_s)e^{-\GH t},  
\end{align}
\noindent where $\phi_s$ is the (\CP-violating) relative weak phase between the $\Bsb$ mixing and $b\to \ccbar s$ decay amplitudes.

The untagged decay rate in Eq.~\ref{eq:decayrate1} provides a probe of $\phi_s$, $\GL$ and $\GH$
in a way that is complementary to direct determinations using \CP violating asymmetries~\cite{LHCb-PAPER-2013-002}.
Approximating Eq.~\ref{eq:decayrate1} by a single exponential 
\begin{align}
\label{eq:decayrate2}
\Gamma_{\bstodsds}(t) + \Gamma_{\Bs\to\Dsp\Dsm}(t)\propto e^{-t/\taueff},
\end{align}
defines the $\bstodsds$ effective lifetime, which can be written as 
$\taueff = \tau_{\Bsb}(1-y_s\cos\phi_s+{\cal{O}}(y_s^2))$~\cite{Fleischer:2011cw,Hartkorn:1999ga}, assuming no direct
\CP violation in the $\bstodsds$ decay.
Here $y_s\equiv\DGs/(2\Gs)$, $\DGs\equiv\GL-\GH$ and $\Gs=(\GH+\GL)/2=1/\tau_{\Bsb}$, where $\tau_{\Bsb}$ is the
flavor-specific $\Bsb$ lifetime.
Using the measured value of $\phi_s=0.01\pm0.07\pm0.01$ rad~\cite{LHCb-PAPER-2013-002}, 
which is in good agreement with the SM expectation of $-0.0363^{+0.0016}_{-0.0015}$~rad~\cite{Charles:2011va}, 
it follows that $\taueff\simeq\GL^{-1}$. 

The most precise measurement to date of the effective lifetime in a \CP-even final state used
$\Bsb\to\Kp\Km$~\cite{LHCb-PAPER-2012-013} decays, and yielded a value $\taueffkk=1.455\pm0.046\stat\pm0.006\syst$~ps.
Loop contributions, both within, and possibly beyond the SM, are expected to be significantly larger in $\Bsb\to\Kp\Km$ than
in $\bstodsds$. These contributions give rise to direct \CP violation in the $\Bsb\to\Kp\Km$ decay~\cite{LHCb-PAPER-2013-040}, 
which lead to differences between $\tau^{\rm eff}$ in these two \CP final state decays,making a comparison of their effective 
lifetimes interesting.
Measurements have also been made in \CP-odd modes, such as $\Bsb\to\jpsi f_0(980)$~\cite{LHCb-PAPER-2012-017,Aaltonen:2011nk} and 
$\Bsb\to\jpsi\KS$~\cite{LHCb-PAPER-2013-015}. The most precise value is from the former, yielding 
$\tau^{\rm eff}_{\Bsb\to\jpsi f_0(980)}=1.700\pm0.040\stat\pm0.026\syst$~ps~\cite{LHCb-PAPER-2012-017}. 
Constraints from these measurements on the (\DGs, $\phi_s$) parameter space
are given in Refs.~\cite{Fleischer:2011cw,Knegjens:2012vc}. Improved precision on the effective lifetimes will enable
more stringent tests of the consistency between the direct measurements of \DGs and $\phi_s$,
and those inferred using effective lifetimes.

In this Letter, the $\bstodsds$ time-dependent decay rate is normalized to the corresponding rate in the $\btodzds$ decay, which has similar 
final state topology and kinematic properties, and a precisely measured lifetime of $\tau_{\Bm}=1.641\pm0.008$~ps~\cite{PDG2012}. As a  result,
many of the systematic uncertainties cancel in the measured ratio. The relative rate is then given by
\begin{align}
\label{eq:decayrate3}
\frac{\Gamma_{\bstodsds}(t) + \Gamma_{\Bs\to\Dsp\Dsm}(t)}{ \Gamma_{\btodzds}(t) + \Gamma_{\Bp\to\Dzb\Dsp}(t)}\propto e^{-\alpha_{su}t}, 
\end{align}
\noindent where $\alpha_{su}=1/\taueff - 1/\tau_{\Bm}$.
A measurement of $\alpha_{su}$ therefore determines $\taueff$. 

The $\Bsb$ meson lifetime is also measured using the flavor-specific, Cabibbo-suppressed $\bstodsd$ decay. 
Its time-dependent rate is normalized to that of the $\btodsd$ decay. In what follows,
the symbol $B$ without a flavor designation refers to either a $\Bm$, $\Bzb$ or $\Bsb$ meson, and $D$ 
refers to either a $\Dz$, $\Dp$ or $\Dsp$ meson. Unless otherwise indicated, charge conjugate final states are included.

The measurements presented use a proton-proton ($pp$) collision data sample corresponding 
to 3~\invfb of integrated luminosity, 1\invfb recorded at a center-of-mass energy of 7\tev and 2\invfb at 8\tev, collected by the 
LHCb experiment.  The \lhcb detector~\cite{Alves:2008zz} is a single-arm forward
spectrometer covering the \mbox{pseudorapidity} range $2<\eta <5$,
designed for the study of particles containing \bquark or \cquark
quarks. The detector includes a high-precision tracking system
consisting of a silicon-strip vertex detector surrounding the $pp$
interaction region, a large-area silicon-strip detector located
upstream of a dipole magnet with a bending power of about
$4{\rm\,Tm}$, and three stations of silicon-strip detectors and straw
drift tubes placed downstream.
The combined tracking system provides a momentum measurement with
relative uncertainty that varies from 0.4\% at 5\gevc to 0.6\% at 100\gevc,
and impact parameter (IP) resolution of 20\mum for
tracks with large transverse momentum (\pt). Ring-imaging Cherenkov detectors~\cite{LHCb-DP-2012-003}
are used to distinguish charged hadrons, and photon, electron and
hadron candidates are identified by a calorimeter system consisting of
scintillating-pad and preshower detectors, an electromagnetic
calorimeter and a hadronic calorimeter.
Muons are identified by a system composed of alternating layers of iron and multiwire
proportional chambers~\cite{LHCb-DP-2012-002}.

The trigger~\cite{LHCb-DP-2012-004} consists of a
hardware stage, based on information from the calorimeter and muon
systems, followed by a software stage, which applies a full event
reconstruction~\cite{LHCb-DP-2012-004,BBDT}. No specific requirement is
made on the hardware trigger decision. Of the $B$ meson candidates
considered in this analysis, about 60\% are triggered at the hardware 
level by one or more of the final state particles in the signal $B$ decay. The remaining 40\% are triggered 
due to other activity in the event.
The software trigger requires a two-, three- or four-track
secondary vertex with a large sum of the transverse momentum of
the tracks and a significant displacement from the primary $pp$
interaction vertices~(PVs). At least one track should have $\pt >
1.7\gevc$ and \chisqip with respect to any
primary interaction greater than 16, where \chisqip is defined as the
difference in \chisq of a given PV reconstructed with and
without the considered particle included. The signal candidates used in this
analysis are required to pass a multivariate software trigger
selection algorithm~\cite{BBDT}.

Proton-proton collisions are simulated using
\pythia~\cite{Sjostrand:2006za,*Sjostrand:2007gs} with a specific \lhcb
configuration~\cite{LHCb-PROC-2010-056}.  Decays of hadronic particles
are described by \evtgen~\cite{Lange:2001uf}, in which final state
radiation is generated using \photos~\cite{Golonka:2005pn}. The
interaction of the generated particles with the detector and its
response are implemented using the \geant
toolkit~\cite{Allison:2006ve, *Agostinelli:2002hh} as described in
Ref.~\cite{LHCb-PROC-2011-006}.

Signal $\bstodsds$ candidates are reconstructed using four final states: (i) $\Dsp\to\Kp\Km\pip,~\Dsm\to\Km\Kp\pim$,
(ii) $\Dsp\to\Kp\Km\pip,~\Dsm\to\pim\pip\pim$, (iii) $\Dsp\to\Kp\Km\pip, \Dsm\to\Km\pip\pim$, and
(iv) $\Dsp\to\pip\pim\pip, \Dsm\to\pim\pip\pim$. In the normalization mode, $\btodzds$, only the 
final state $\Dz\to\Km\pip,~\Dsm\to\Km\Kp\pim$ is used. For the $\bstodsd$ decay and the corresponding
$\Bz$ normalization mode, the $\Dm\to\Kp\pim\pim,~\Dsp\to\Kp\Km\pip$ final state is used.
Loose particle identification (PID) requirements are
imposed on kaon and pion candidates, with efficiencies typically in excess of 95\%. 
The $D$ candidates are required to have masses within
25\mevcc of their known values~\cite{PDG2012} and to have vertex separation from the $B$ vertex
satisfying $\chisqvs>2$. Here $\chisqvs$ is the increase in $\chi^2$ of the parent ($B$) vertex fit when the ($D$ meson) 
decay products are constrained to come from the parent vertex, relative to the nominal fit. 
To suppress the large background from $\bstodspipipi$ decays, $\Dsm\to\pim\pip\pim$ candidates are
required to have $\chisqvs>6$. As the signatures of $b$-hadron decays to double-charm final states are similar, 
vetoes are employed to suppress the cross-feed resulting from particle misidentification, following
Ref.~\cite{LHCb-PAPER-2012-050}. For the $\Dsp\to\Kp\pim\pip$ decay, an additional veto to suppress
cross-feed from $\Dp\to\Km\pip\pip$ with double-misidentification is employed, which renders this background
negligible. Potential
background to $\Dsp$ decays from $\Dstarp\to\Dz\pip$ with $\Dz\to\Kp\Km,~\pip\pim$ is also removed by 
requiring the mass difference, $M(\Dz\pip)-M(\Dz)>150$\mevcc. 
The production point of each $B$ candidate is taken as the PV with the smallest $\chisqip$ value. 
All $B$ candidates are refit taking both $D$ mass and vertex constraints into account~\cite{Hulsbergen:2005pu}.

The efficiencies of the PID and veto requirements are evaluated using dedicated $\Dstarp\to\Dz\pip,~\Dz\to\Km\pip$ calibration
samples collected at the same time as the data. The kinematic distributions of kaons and pions from the calibration sample
are reweighted using simulation to match those of the $B$ decays under study. 
The combined PID and veto efficiencies are 91.4\% for $\btodzds$, 88.0\% for $(\Bsb,~\Bz)\to\Dm\Dsp$, 
and 86.5\%, 90.8\%, 86.6\%, and 95.9\% for the $\bstodsds$ final states (i)$-$(iv), respectively.

To further improve the signal-to-background ratio, a boosted decision tree (BDT)~\cite{Breiman,AdaBoost} algorithm
using seventeen input variables
is employed. Five variables from the $B$ candidate are used, including $\chisqip$, the vertex fit $\chisqvtx$ (with
$D$ mass, and vertex constraints), the PV $\chisqvs$, \pt, and a \pt asymmetry variable~\cite{LHCb-PAPER-2012-055}.
For each $D$ daughter, $\chisqip$, the flight distance from the $B$ vertex normalized by its uncertainty, and
the maximum distance between the trajectories of any pair of particles in the $D$ decay, are used. Lastly, for each $D$ 
candidate, the minimum \pt, and both the smallest and largest $\chisqip$, among the $D$ daughter particles are used.
The BDT uses simulated decays to emulate the signal and wrong-charge final states from data with masses larger than
5.2\gevcc for the background. Here, wrong-charge refers to $D_s^{\pm}D_s^{\pm}$, $D^{\pm}D_s^{\pm}$, and $\Dz\Dsp$ combinations, 
where in the latter case we remove candidates within 30\mevcc of the $\Bp$ mass~\cite{PDG2012}, to remove the
small doubly-Cabibbo-suppressed decay contribution to this final state. 
The selection requirement on the BDT output is chosen to maximize the expected $\bstodsds$ signal significance,
corresponding to signal and background efficiencies of about 97\% and 33\%, respectively.
More than one candidate per event is allowed, but after all selections the fraction of events with multiple candidates
is below 0.25\% for all modes. 

\begin{figure}[tb]
\centering
\includegraphics[width=0.65\textwidth]{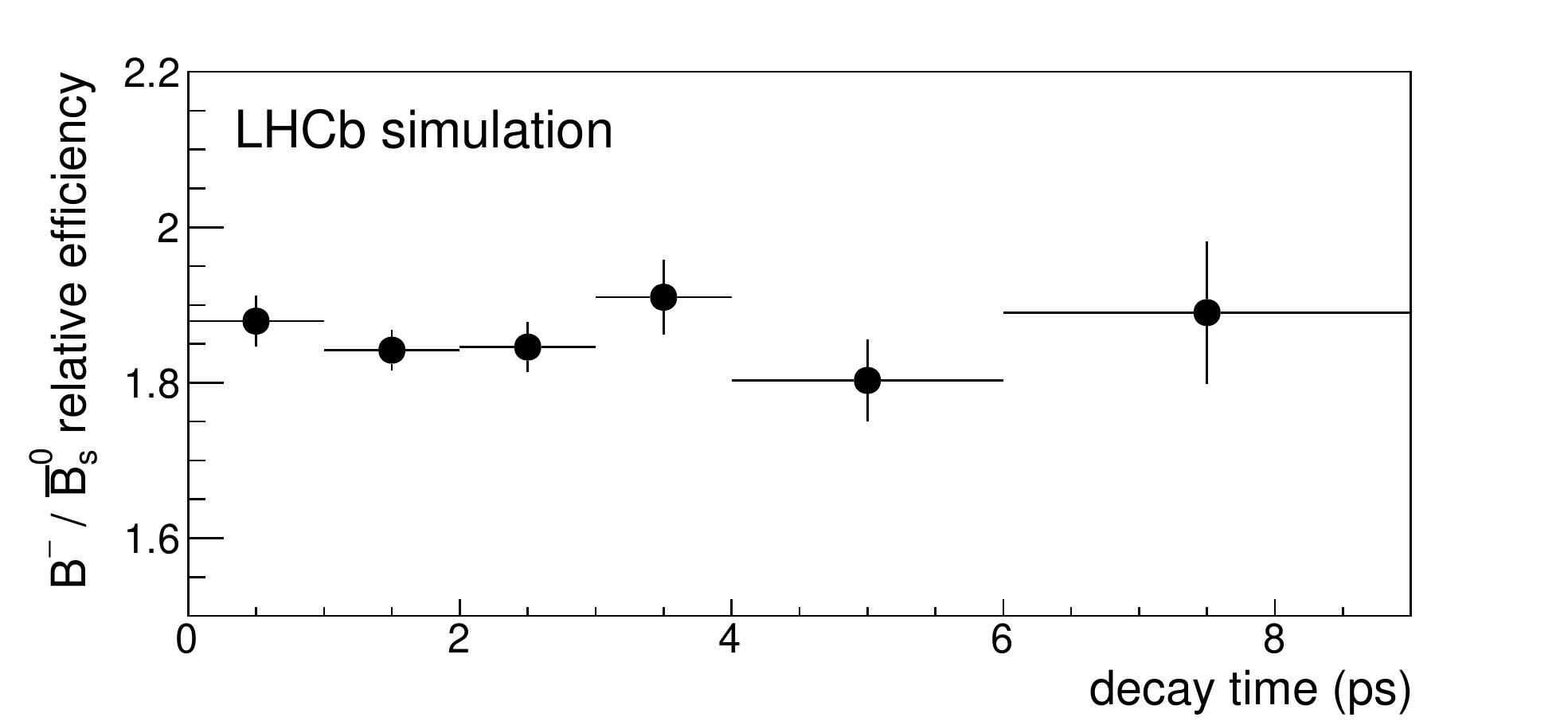}
\caption{Ratio of selection efficiencies for $\btodzds$ relative to $\bstodsds$ decays
as a function of decay time. The uncertainties shown are due to finite simulated sample sizes.}
\label{fig:relEff1}
\end{figure}

For the lifetime analysis, we consider only $B$ candidates with reconstructed decay time less than 9~ps.
Signal efficiencies as functions of decay time are determined using simulated decays after all selections, except those that 
involve PID, as described above. The resulting $\Bm$ to $\Bsb$ relative efficiency as a function of decay time
is shown in Fig.~\ref{fig:relEff1}, where six decay time bins with widths ranging between 1 and 3~ps are used.
For the $\bstodsds$ decay, the efficiency used in the ratio is the weighted average of the $\Dsp\Dsm$ final states (i)$-$(iv), 
where the weights are obtained from the observed yields in data.
The efficiency accounts for the migration between bins, which is small since the resolution on the reconstructed time of
$\sim$50~fs is much less than the bin width. Moreover, the time resolution is nearly identical for the signal and normalization modes,
and is independent of the reconstructed lifetime.
The relative efficiency is consistent with being independent of decay time, however, the computed bin-by-bin efficiencies are used to correct 
the data.

\begin{figure}[tb]
\centering
\includegraphics[width=0.49\textwidth]{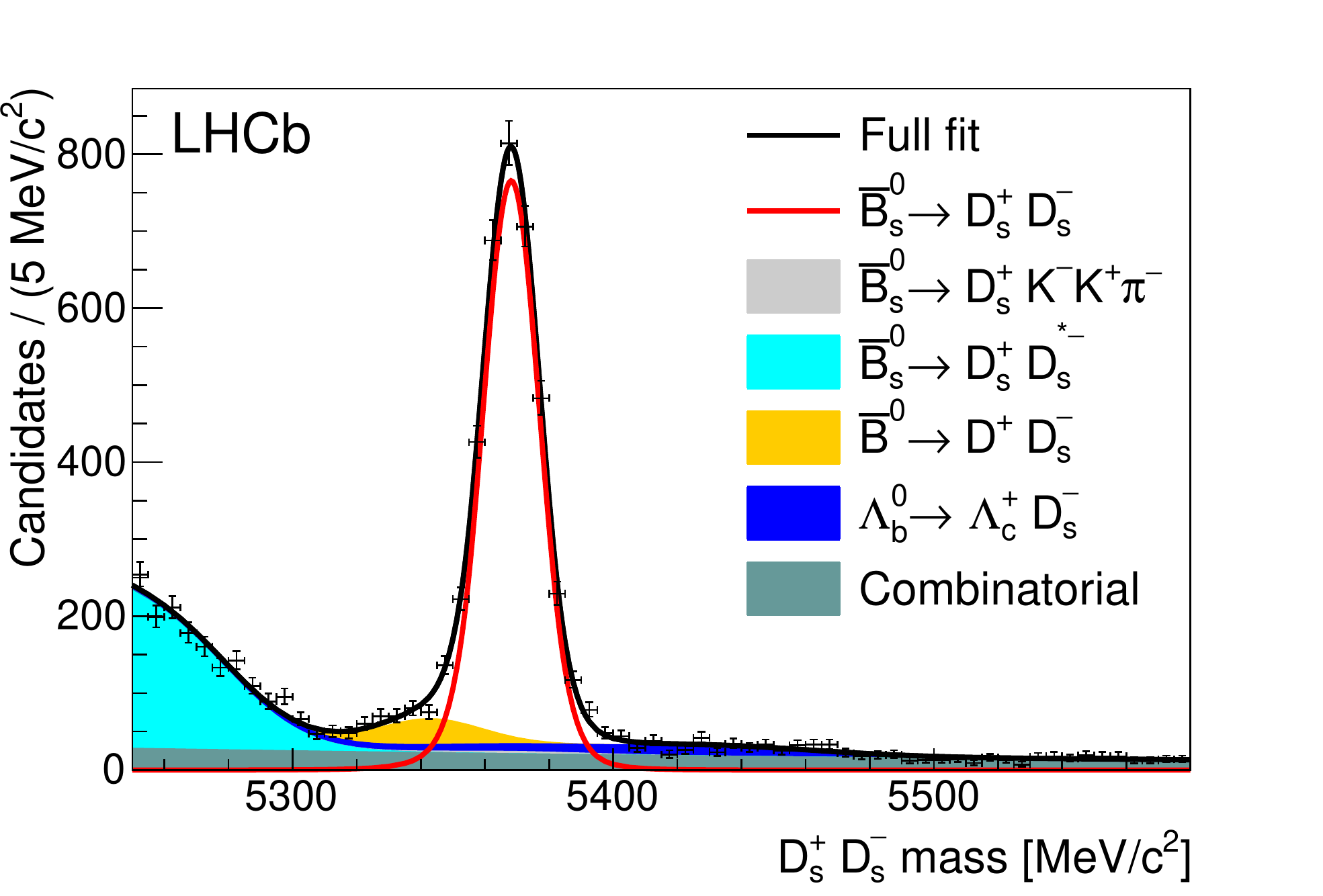}
\includegraphics[width=0.49\textwidth]{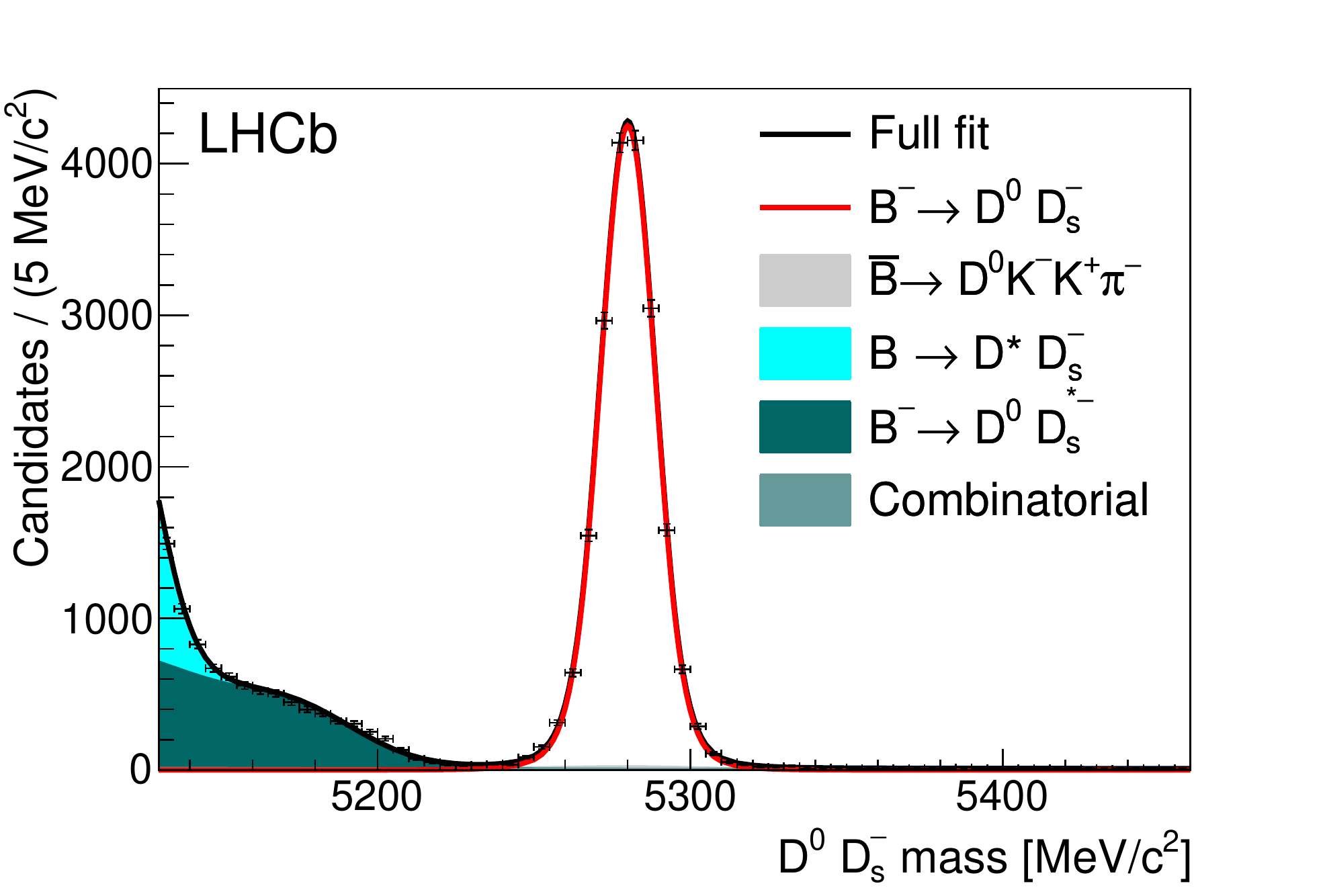}
\caption{\small{Mass distributions and fits to the full data sample for (left) $\bstodsds$ and (right) $\btodzds$
candidates. The points are the data and the curves and shaded regions show
the fit components. }}
\label{fig:fitFullSample}
\end{figure}

The mass distributions for the signal, summed over the four final states, and the normalization modes 
are shown in Fig.~\ref{fig:fitFullSample}, along
with the results of binned maximum likelihood fits. The $B$ signal shapes are each
modeled using the sum of two Crystal Ball (CB) functions~\cite{Skwarnicki:1986xj} with a common mean.
The shape parameters are fixed from fits to simulated signal decays, with the exception of the resolution parameter, 
which is found to be about 15\% larger in data than simulation.
The shape of the low-mass background from partially reconstructed decays, where either a photon or pion is missing, is 
obtained from simulated decays, as are the cross-feed background shapes from $\btodsd$ and $\Lb\to\Lc\Dsm$ decays ($\bstodsds$ channel only).
An additional peaking background due to $B\to D\Km\Kp\pim$ decays is also included in the fit. 
Its shape is obtained from simulation and the yield is fixed to be 1\% of the signal yield from a
fit to the $D$ mass sidebands. The combinatorial background shape is described by
an exponential function with the shape parameter fixed to the value obtained from a fit to the mass spectrum
of wrong-charge candidates.
All yields, except that of the $B\to D\Km\Kp\pim$, are freely varied in the fit to the full data sample. 

\begin{figure}[tb]
\centering
\includegraphics[width=0.65\textwidth]{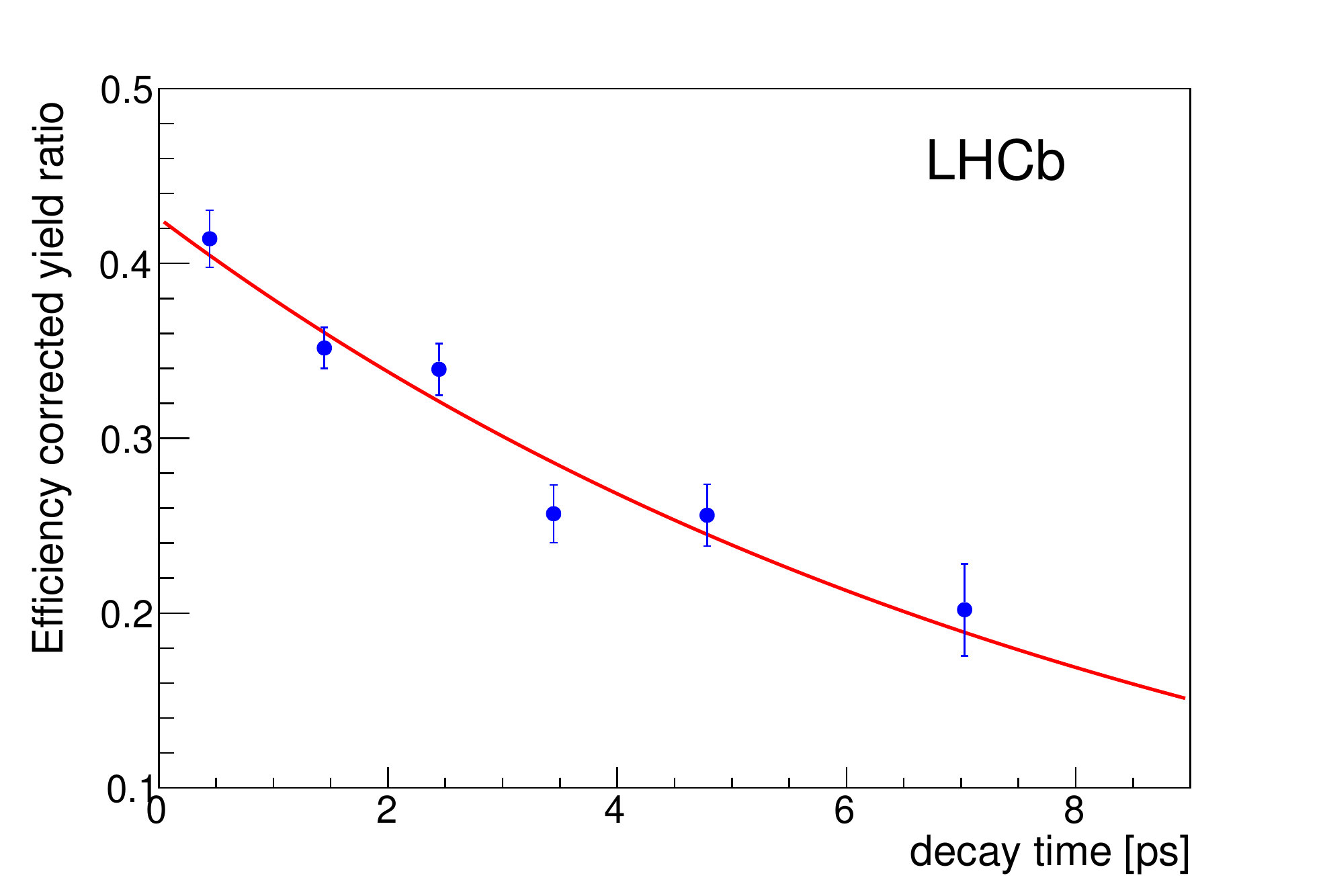}
\caption{\small{Efficiency corrected yield ratio of $\bstodsds$ relative to $\btodzds$ as a function of decay time, 
along with the exponential fit. The uncertainties are statistical only.}}
\label{fig:CorrYieldRatio_BsOverB_nominal}
\end{figure}

In total, we observe 3499\,$\pm$\,65 $\bstodsds$ and 19,432\,$\pm$\,140 $\btodzds$ decays.
The data are split into the time bins shown in Fig.~\ref{fig:relEff1}, and each mass distribution is fitted with
the CB widths fixed to the values obtained from the full fit. The independence of the signal shape parameters on 
decay time is validated using simulated decays. The ratios of yields are then computed, and corrected by the
relative efficiencies shown in Fig.~\ref{fig:relEff1}. Figure~\ref{fig:CorrYieldRatio_BsOverB_nominal} shows the
efficiency-corrected yield ratios as a function of decay time. The data points are placed at the average
time within each bin assuming an exponential form $e^{-t/(1.5\,{\rm ps})}$. Fitting an exponential function to the 
data yields the result $\alpha_{su}=0.1156\pm0.0139$~ps$^{-1}$. The uncertainty in the fitted slope
due to using the value of 1.5~ps to get the average time in each bin is negligible. Using the known $\Bm$ lifetime, 
$\taueff$ is determined to be $1.379\pm0.026\stat$~ps.

As a cross-check, the full analysis is applied to the $\btodzds$ and $\btodsd$ decays, treating the
former as the signal mode and the latter as the normalization mode. The fitted value for
$\alpha\equiv 1/\tau_{\Bz}-1/\tau_{\Bm}$ is $0.0500\pm0.0076$~ps$^{-1}$, in excellent agreement with
the expected value of $0.0489\pm0.0042$~\cite{PDG2012}. 
This check indicates that the relative lifetime measurements are insensitive to small differences in the
number of charged particles or lifetimes of the $D$ mesons in the final state. The $\btodsd$ mode could have also 
been used as
a normalization mode for the $\bstodsds$ time-dependent rate measurement, but due to limited simulated sample sizes
it would have led to a larger systematic uncertainty.

As the method for determining $\taueff$ relies on ratios of yields and efficiencies, many systematic
uncertainties cancel. The robustness of the relative acceptance is tested by subdividing the sample
into mutually exclusive subsamples based on (i) center of mass energy, (ii) $\Dsm\Dsp$ final states, and 
(iii) hardware trigger decision, and searching for deviations larger than 
those expected from the finite sizes of the samples. The results from all checks were found to be within one standard deviation
of the average. Based on the largest deviation, we assign a 0.010~ps systematic uncertainty
due to the modeling of the relative acceptance. The statistical precision on the relative acceptance, as obtained
from simulation, contributes an uncertainty of 0.011~ps. 
Using a different signal shape to fit the data leads to 0.003~ps uncertainty. If the combinatorial background
shape parameter is allowed to freely vary in each time bin fit, we find a deviation of
0.001~ps from the nominal value of $\taueff$, which is assigned as a systematic uncertainty. 
Due to the presence of a non-trivial acceptance function, the 
result of fitting a single exponential to the untagged $\Bs$ decay time distribution does not coincide precisely  
with the formal definition of the effective lifetime~\cite{DeBruyn:2012wj}. The deviation between
$\taueff$ and the single exponential fit is at most 0.001~ps~\cite{DeBruyn:2012wj}, which is assigned
as a systematic uncertainty. The precision on the $\Bm$ lifetime leads to 0.008~ps uncertainty
on the value of $\taueff$. Summing these deviations in quadrature, we obtain a total systematic uncertainty of 0.017~ps.
In converting to a measurement of $\GL$, an additional uncertainty due to a small \CP-odd component
of expected size $1-\cos\phi_s=(0.1\pm3.2)\times10^{-3}$~\cite{LHCb-PAPER-2013-002} leads to a bias no larger than 
$-0.001$~ps$^{-1}$. This is included in the $\GL$ systematic uncertainty.

The value of $\taueff$ and the corresponding decay width of the light $\Bsb$ mass eigenstate are determined to be
\begin{align*}
\taueff &= 1.379\pm0.026\pm0.017~{\rm ps}, \\
\GL &= 0.725\pm0.014\pm0.009~{\rm ps}^{-1},
\end{align*}
where the first uncertainty is statistical and the second is systematic. 
These are the first such measurements using the $\bstodsds$ decay.
The measured effective lifetime represents the most precise measurement of the width of the light $\Bsb$ mass eigenstate, and
is about one standard deviation lower than the value obtained using $\Bsb\to\Kp\Km$ decays~\cite{LHCb-PAPER-2012-013}.
Compared to the $\bstodsds$ decay, which is
dominated by tree-level processes, the $\Bsb\to\Kp\Km$ decay is expected to have larger relative contributions from SM-loop 
amplitudes~\cite{Fleischer:1999pa,Fleischer:2007zn,Fleischer:2011cw}, 
and therefore one should not naively average the effective lifetimes from these two decays. Moreover, if non-SM particles contribute 
additional amplitudes, 
their effect is likely to be larger in $\Bsb\to\Kp\Km$ than in $\bstodsds$ decays~\cite{Fleischer:2010ib}. 

The value of $\GL$ obtained in this analysis may be
compared to the value inferred from the time-dependent analyses of $\jpsi\Kp\Km$ and $\jpsi\pip\pim$ decays. Using the
values $\Gs=0.661\pm0.004\pm0.006$~ps$^{-1}$ and $\DGs=0.106\pm0.011\pm0.007$~ps$^{-1}$~\cite{LHCb-PAPER-2013-002}, 
we find $\GL=0.714\pm0.010$~ps$^{-1}$, in good agreement with the value obtained from $\taueff$.

\begin{figure}[tb]
\centering
\includegraphics[width=0.65\textwidth]{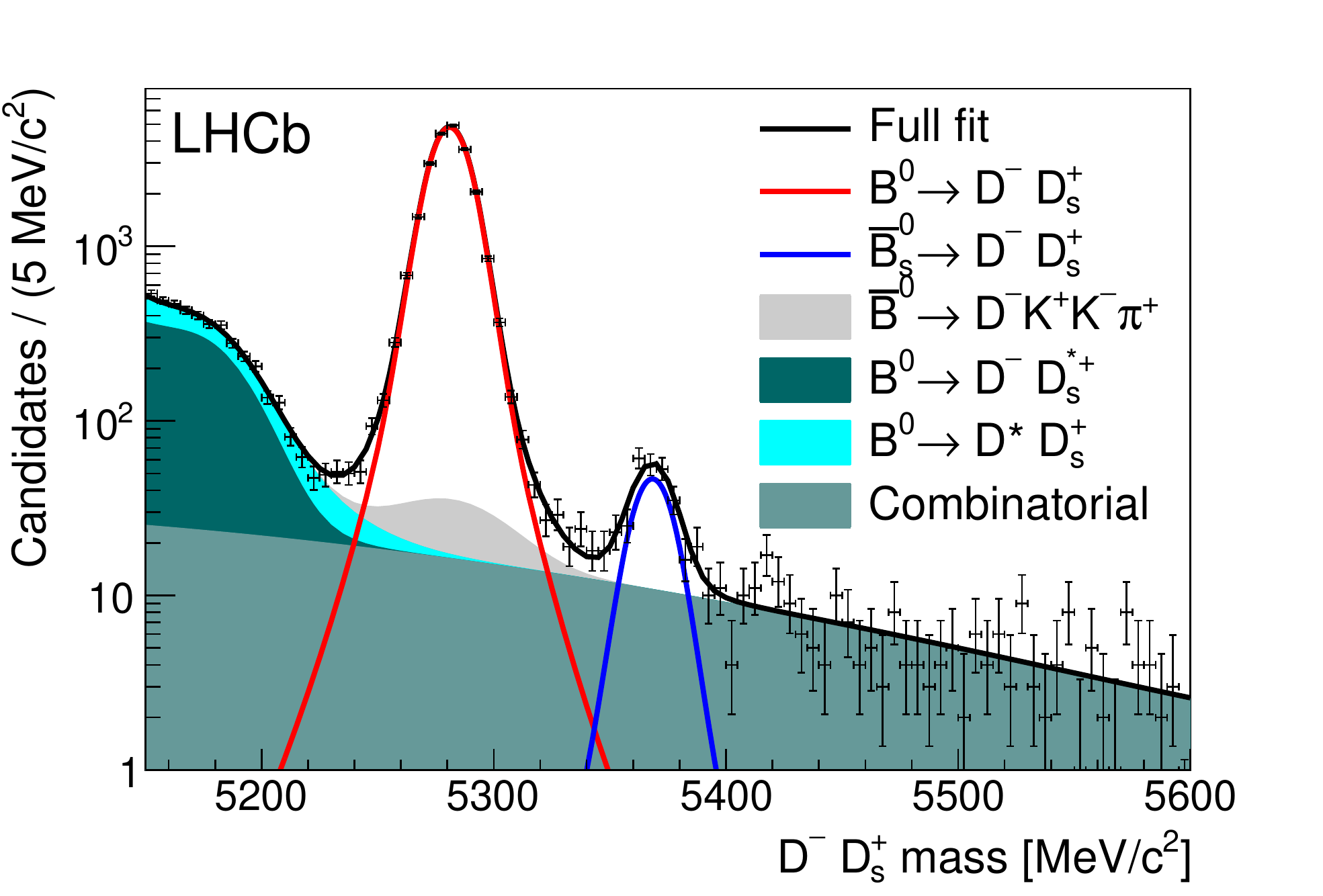}
\caption{\small{Mass distribution and fits to the full data sample for $\Bsb$ and $\Bz$ decays into the 
$\Dm\Dsp$ final state. The points are the data and the curves and shaded regions show the fit components.}}
\label{fig:fitBs2DsD}
\end{figure}

The effective lifetime of the flavor-specific $\bstodsd$ decay is also measured, using the $\btodsd$ decay for
normalization.
The technique is identical to that described above, with the simplification that the relative efficiency equals 
one, since the final states are identical. Effects due to the mass difference between the $\Bsb$ and $\Bz$ mesons are 
negligible. A tighter BDT selection is imposed to optimize the expected signal-to-background ratio, which results in signal
and background efficiencies of 87\% and 11\%, respectively. The mass spectrum and the corresponding fit are 
shown in  Fig.~\ref{fig:fitBs2DsD}, where the fitted components are analogous to those described previously.
A total of $230\pm18$ $\bstodsd$ and 21,195\,$\pm$\,147 $\btodsd$ decays are obtained.
The time bins are the same as above, except the 6$-$9 ps bin is dropped, since the yield in the signal mode beyond 6~ps is negligible.
The relative decay rate is fitted to an exponential form ${\cal{C}}e^{-\beta t}$, where ${\cal{C}}$ is a normalization constant.
The fitted value of $\beta$ is $0.000\pm0.068$~ps$^{-1}$.
The systematic uncertainty due to the signal shape is 0.007~\ps, obtained by using a different signal shape function. The
exponential background shape is fixed in the nominal fit using $D^{\pm}D_s^{\pm}$ candidates, and a systematic uncertainty of
0.010~ps is determined by allowing its shape parameter to vary freely in the fit. In determining the effective lifetime, 
an uncertainty of 0.007~\ps due to the limited precision of the $\Bz$ lifetime~\cite{PDG2012} is also included. The resulting 
effective lifetime in the $\bstodsd$ mode is
\begin{align*}
\tau^{\rm eff}_{\bstodsd} = 1.52\pm0.15\pm0.01~{\rm ps}.
\end{align*}
This is the first measurement of the $\Bsb$ lifetime using the $\bstodsd$ decay. Its value is consistent with previous direct and
indirect measurements of the $\Bsb$ lifetime in other flavor-specific decays.

In summary, we report the first measurement of the $\bstodsds$ effective lifetime and present the most precise direct measurement
of the width of the light $B_s$ mass eigenstate. Their values are $\taueff = 1.379\pm0.026\pm0.017~{\rm ps}$ and
 $\GL=0.725\pm0.014\pm0.009~{\rm ps}^{-1}$. The $\GL$ result is consistent with the value
obtained from previously measured values of $\DGs$ and $\Gs$~\cite{LHCb-PAPER-2013-002}. We also determine the average $\Bsb$ 
lifetime to be $1.52\pm0.15\pm0.01~{\rm ps}$ using the $\bstodsd$ decay, which is consistent with other measurements.

%% file: acknowledgements.tex
\section*{Acknowledgements}

\noindent We express our gratitude to our colleagues in the CERN
accelerator departments for the excellent performance of the LHC. We
thank the technical and administrative staff at the LHCb
institutes. We acknowledge support from CERN and from the national
agencies: CAPES, CNPq, FAPERJ and FINEP (Brazil); NSFC (China);
CNRS/IN2P3 and Region Auvergne (France); BMBF, DFG, HGF and MPG
(Germany); SFI (Ireland); INFN (Italy); FOM and NWO (The Netherlands);
SCSR (Poland); MEN/IFA (Romania); MinES, Rosatom, RFBR and NRC
``Kurchatov Institute'' (Russia); MinECo, XuntaGal and GENCAT (Spain);
SNSF and SER (Switzerland); NAS Ukraine (Ukraine); STFC (United
Kingdom); NSF (USA). We also acknowledge the support received from the
ERC under FP7. The Tier1 computing centres are supported by IN2P3
(France), KIT and BMBF (Germany), INFN (Italy), NWO and SURF (The
Netherlands), PIC (Spain), GridPP (United Kingdom). We are thankful
for the computing resources put at our disposal by Yandex LLC
(Russia), as well as to the communities behind the multiple open
source software packages that we depend on.